\documentclass[twocolumn]{article}
\usepackage{geometry}
\usepackage{textcomp}
\usepackage{caption}
\usepackage{graphicx}
\usepackage{amsmath}
\usepackage{amssymb}
\usepackage{textcomp}
\usepackage[dvipsnames]{xcolor}
\usepackage{authblk}
\usepackage{datetime}
\usepackage{gensymb}
\usepackage{wrapfig}
\usepackage{booktabs}
\usepackage{ulem}
\usepackage[numbers]{natbib}
\usepackage{hyperref}
\hypersetup{
    citecolor = blue,
    filecolor = black,
    urlcolor = blue,
    colorlinks = true, %set true if you want colored links
    linktoc = all,     %set to all if you want both sections and subsections linked
    linkcolor = blue,  %choose some color if you want links to stand out
}

\newcommand{\onlinecite}[1]{\hspace{-1 ex} \nocite{#1}\citenum{#1}} 
\definecolor{lightgrey}{gray}{0.9}
\newcommand{\code}[1]{\texttt{#1}}
  
\title{Programmable Superconducting Optoelectronic Single-Photon Synapses with Integrated Multi-State Memory}

\author[1]{ Bryce A. Primavera$^{1,2}$, \large{Saeed Khan$^{1,2}$, Richard P. Mirin$^1$, Sae Woo Nam$^1$, and Jeffrey M. Shainline$^{1*}$}
\\
\vspace{0.5em}
\textit{\large{$^1$National Institute of Standards and Technology}}
\\
\textit{\large{325 Broadway, Boulder, CO, USA, 80305}}
\vspace{0.5em}
\vspace{0.5em}\\
\textit{\large{$^2$Department of Physics}}\\
\textit{\large{University of Colorado Boulder}}\\
\textit{\large{390 UCB, Boulder, CO, USA, 80309}}\\
\vspace{0.5em}
\small{$^*$jeffrey.shainline@nist.gov}
}
\date{\today}%\today

\begin{document}

\twocolumn[
\begin{@twocolumnfalse}
\maketitle
\begin{abstract}
The co-location of memory and processing is a core principle of neuromorphic computing. A local memory device for synaptic weight storage has long been recognized as an enabling element for large-scale, high-performance neuromorphic hardware. In this work, we demonstrate programmable superconducting synapses with integrated memories for use in superconducting optoelectronic neural systems. Superconducting nanowire single-photon detectors and Josephson junctions are combined into programmable synaptic circuits that exhibit single-photon sensitivity, memory cells with more than 400 internal states, leaky integration of input spike events, and 0.4\,fJ programming energies (including cooling power). These results are attractive for implementing a variety of supervised and unsupervised learning algorithms and lay the foundation for a new hardware platform optimized for large-scale spiking network accelerators.

\vspace{3em}
\end{abstract}
\end{@twocolumnfalse}
]

\setcounter{tocdepth}{1}
\setcounter{secnumdepth}{4}
%\tableofcontents

\section{\label{sec:introduction}Introduction}
Computing performance has been limited by the von Neumann bottleneck for decades \cite{sebastian2020memory}. These memory access challenges, in conjunction with the rise of memory-intensive deep learning applications, have led to a reexamination of computing architecture in recent years. Neuromorphic architectures modelled after biological neural systems are candidates for the next generation of artificial intelligence hardware. Computational and architectural motifs such as distributed analog computation, highly interconnected communication networks, and co-location of memory and information processing, are key to the impressive performance of biological neural systems. These principles can serve as broad guidelines for hardware engineers.

Superconducting optoelectronic networks (SOENs) were introduced to maximize scalability while adhering to such biologically-derived principles \cite{shainline2019superconducting, shainline2021optoelectronic}. With this hardware, high-speed, low-power processing is performed with superconducting analog spiking neural circuits based on Josephson junctions (JJs). These superconducting neurons are embedded in a highly interconnected optical network that enables direct communication between each neuron and thousands of downstream synapses. Spiking events are encoded as few-photon pulses of light that are directly transmitted between an integrated light source at each neuron and single-photon-sensitive detectors at each synaptic connection. This single-photon sensitivity manifests itself in the extreme fan-out capability of superconducting optoelectronic neurons by placing the physically minimal performance requirements on the light sources. Direct synaptic connections ensure communication latency is independent of network scale and activity up to systems of billions of neurons \cite{shainline2019superconducting}.

SOENs synapses were realized in Ref.\onlinecite{khan2022superconducting}, enabled by the monolithic integration of superconducting-nanowire single-photon detectors (SPDs) with JJs. While those synapses demonstrated single-photon sensitivity and biologically relevant computations like leaky integration and tunable synaptic weights, the synaptic weights were defined with current biases generated off-chip. The requirement of an independent current source for each synaptic weight is an unscalable solution that contradicts the principle of co-location of processing and memory.

Like other neuromorphic platforms, SOENs stands to greatly benefit from a local, multi-state memory that can be programmed for hardware-in-the-loop training or updated based on network activity for on-chip learning. While room-temperature synaptic memory technologies remain an intensely active research area centered on materials development and integration (be it memristive, ferroelectric, or phase-change materials) \cite{burr2017neuromorphic}, suitable superconducting memories are a decades-old technology that require no changes to standard superconducting fabrication processes \cite{crowe1957trapped, tahara1987vortex}. These ``superconducting loop memories'' store information as circulating currents trapped in superconducting loops. With identically zero resistance in the loop, the circulating current persists indefinitely. Further, such memories permit high bit-depths, low programming energy, high endurance, and programming pulses easily produced by integrated JJ circuitry \cite{schegolev2020learning, primavera2021considerations}. In this work, we adapt the synapses of Ref.\,\onlinecite{khan2022superconducting} for integration with superconducting loop memory to demonstrate programmable single-photon sensitive optoelectronic synapses. 

\begin{figure*}[!ht]
    \centering
    \includegraphics{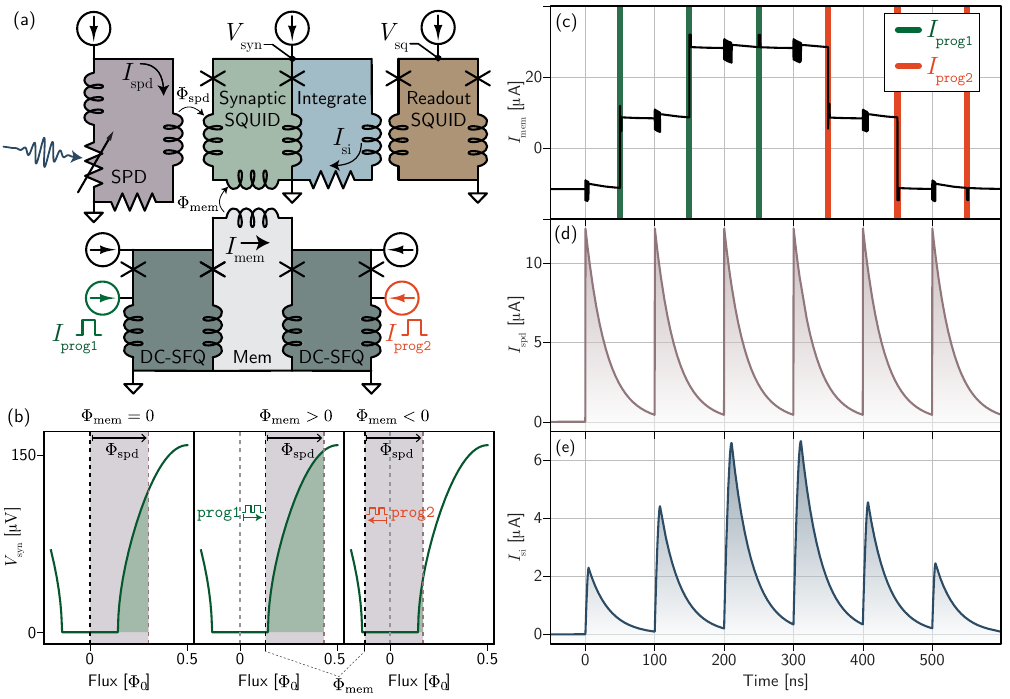}
    \caption{Synapse concept and simulation. (a) Circuit diagram. Current $I_\mathrm{si}$ is added to the integration loop (blue) each time the SPD (purple) detects a photon. The amount of $I_\mathrm{si}$ produced is mediated by the synaptic SQUID (light green). A programmable memory loop (dark green and tan blocks) determines the quiescent point of the synaptic SQUID, which establishes the synaptic weight. The voltage across the readout SQUID ($V_\mathrm{sq}$) is measured as a proxy for $I_\mathrm{si}$ in the experiments. (b) A section of the synaptic SQUID's flux-voltage transfer function is shown in green. The SQUID is initially biased below threshold. When the SPD detects a photon, the flux $\Phi_\mathrm{spd}$ drives the SQUID into the active region of its response. Changing the state of the memory loop corresponds to shifting the initial flux penetrating the SQUID ($\Phi_\mathrm{mem}$). Higher values of $\Phi_\mathrm{mem}$ result in the SPD driving the SQUID further into the voltage state and more current added to the integration loop. (c) Simulation of the stored current ($I_\mathrm{mem}$) in a 3-state memory loop. Unless the loop is saturated, each \code{prog1} pulse adds one fluxon to the loop and each \code{prog2} pulse subtracts one fluxon from the loop. (d) The current $I_\mathrm{spd}$ diverted away from the SPD each time a photon is detected. This waveform is determined by the optimal parameters for photon detection and is unaffected by changes to the synaptic weight. (e) The post-synaptic integrated current ($I_\mathrm{si}$) at three different weights for six photon detection events, showcasing the effect of the memory loop. Note that the synaptic decay time is 30\,ns in this simulation in order to present the SPD and integration loop dynamics on similar timescales. In the fabricated devices, the decay time is 6.25\,\textmu s.}
    \label{fig:schematic}
\end{figure*}

\section{\label{sec:circuits}Circuitry}
\subsection{\label{subsec:memory}Superconducting Loop Memory}
The phenomenon of flux quantization in superconducting loops is one of the most well-known manifestations of quantum mechanics at the macroscale \cite{van1998principles}. The requirement that the wavefunction of the superconducting state be single valued ensures that the magnetic flux penetrating the loop must be an integer multiple of the magnetic flux quantum, $\Phi_0$ ($2.07 \times 10^{-15}$\,Wb). Equivalently, it is useful to think of the total flux as being composed of an integer number of flux quanta or fluxons, each carrying $\Phi_0$ of flux. For an uninterrupted superconducting loop, the penetrating flux will for all time remain the same as when superconductivity was first established. However, if JJs are embedded in the loop, the amount of trapped flux can be changed in increments of a single fluxon. This is the basis of superconducting loop memory, where the number of trapped fluxons in the loop is used to define the state of the memory cell. Trapped fluxons are associated with an induced shielding current in the memory loop,
\begin{equation}
    I_\mathrm{mem} = \frac{N_{\phi}\Phi_0}{L_\mathrm{mem}}
\end{equation}
where $N_\phi$ is the number of stored fluxons and $L_\mathrm{mem}$ is the loop inductance. This current will persist indefinitely, resulting in loop memory's exceptional retention times. In digital computing applications it is common to employ a binary memory cell, representing a zero by the absence of trapped flux and a one by the presence of a single fluxon. In the present neural application, a many-state memory is desirable. In this work we demonstrate three memory loops with $N_{\phi} = $ 8, 28, and 415, which we refer to as 3\,bit, 5\,bit, and 8\,bit synapses, respectively.

While there are several variations of circuitry for programming the states of memory loops, we pursue the circuit in the lower part of Fig.\,\ref{fig:schematic}\,(a) for its simplicity. In this case, the memory loop is inserted between two DC-SFQ circuit blocks. These DC-SFQ converters produce exactly one fluxon or single flux quantum (SFQ) each time the input current ($I_\mathrm{\code{\code{prog1}}}$ or $I_\mathrm{\code{\code{prog2}}}$) crosses a threshold set by the DC current bias \cite{van1998principles,kadin1999introduction}. By placing one on either side of the memory loop, fluxons can be added or subtracted one at a time by applying a programming pulse to the appropriate DC-SFQ converter. The operation of the DC-SFQ converters is independent of programming pulse width and largely independent of the precise programming pulse height, as long as the pulse amplitude exceeds the threshold set by the current bias (programming pulse height does affect the saturation level of the memory loop as described in the next section). The number of states available to the loop is ultimately limited by the loop inductance, with larger inductances supporting more states since the stored current per fluxon is inversely proportional to $L_\mathrm{mem}$. In the data section, we demonstrate memories with three different inductances (154\,pH, 620\,pH, and 22.5\,nH) to realize the three different storage capacities.

\subsection{\label{subsec:synapse_cricuits}Synaptic Circuits}
In order to couple memory loops to synapses, the previous generation of synaptic circuits \cite{khan2022superconducting} underwent a major redesign [Fig.\,\ref{fig:schematic}\,(a)]. In the new design, each synapse is based around a superconducting quantum interference device (SQUID) connected to an integration loop. A SQUID is a common device in superconducting electronics in which two Josephson junctions are current-biased in parallel (in this work we refer specifically to the DC SQUID). The SQUID acts as a flux to voltage transducer, where the voltage across the device is a periodic function of the magnetic flux penetrating the loop. For these circuits, we restrict the range of inputs so that only one period of the response plays a role and the SQUID transfer function is a monotonic, nonlinear function of input flux, as shown in Fig.\,\ref{fig:schematic}\,(b). The SPD is inductively coupled to the SQUID such that any photon detection event results in magnetic flux ($\Phi_\mathrm{spd}$) applied to the SQUID. $\Phi_\mathrm{spd}$ is proportional to the current $I_\mathrm{spd}$ and decays with a time constant set by the detector (30\,ns for this work). In this application, we bias the SQUID below the critical current so that it is in a state of zero voltage when no synaptic activity occurs. The flux coupled in by the SPD must exceed a bias-dependent threshold to activate the SQUID. When the SPD detects a photon, the SQUID will be driven into the voltage state where it produces a series of fluxons at a rate proportional to the voltage $V_\mathrm{syn}$. These fluxons are stored as current $I_\mathrm{si}$ in the integration loop (blue). Adding a resistor to this integration loop causes the current to decay exponentially, resulting in leaky integrator behavior. $I_\mathrm{si}$ is analogous to the post-synaptic potential of a biological synapse and is the signal that will be fed into the neuron cell body (or dendritic tree).

The synaptic weighting mechanism demonstrated here functions by adjusting a flux bias to the synaptic SQUID. The memory loop couples flux ($\Phi_\mathrm{mem}$) into the SQUID, acting as an offset flux that can move the SQUID's quiescent point closer or further from its turn-on point. The total number of fluxons generated during the SPD response depends on the total flux coupled in, which is the sum of the contribution from the SPD ($\Phi_\mathrm{spd}$) and the contribution from the memory cell ($\Phi_\mathrm{mem}$). Incoming SPD flux will then drive the synaptic SQUID more or less strongly depending on where $\Phi_\mathrm{mem}$ is placed relative to the turn-on point of the SQUID transfer function [Fig.\,\ref{fig:schematic}\,(b)]. $\Phi_\mathrm{mem}$ is programmed by placing an integer number of fluxons in the memory loop using a sequence of programming pulses to ports \code{prog1} and \code{prog2}.

In Fig.\,\ref{fig:schematic}\,(c-e), we simulate the current in various parts of the synapse in the time domain. In Fig.\,\ref{fig:schematic}\,(c), $I_\mathrm{mem}$ is adjusted one fluxon at a time with a series of programming pulses. Recall $I_\mathrm{mem}$ is inductively coupled to the SQUID and therefore proportional to the flux offset that will ultimately determine the synaptic weight. We simulate a memory loop with only three states to illustrate the phenomenon of saturation. Saturation occurs when the memory loop has railed at either its maximum or minimum level of current. If one of the programming ports is pulsed repeatedly, current will be diverted from the DC-SFQ bias and into $I_\mathrm{mem}$ with each fluxon produced. Eventually there is no longer enough bias current for the DC-SFQ converter to produce a fluxon in response to the next programming pulse, and the current in the memory loop will saturate. We observe positive and negative saturation after the third \code{prog1} pulse and third \code{prog2} pulse respectively. Memory loop saturation is a beneficial behavior, as it is used to keep the synaptic SQUID operating in a useful range of its response. In Fig.\,\ref{fig:schematic}\,(d), the SPD detects a series of six photons. Each of these detection events results in exactly the same amount of diverted current from the SPD, regardless of the state of the synapse. This allows the SPD to be biased for optimal detection efficiency at any synaptic weight and permits many SPDs to be biased in series. In contrast, the amplitude of $I_\mathrm{si}$ per detected photon is a function of the state of the memory loop and can be programmed as desired [Fig.\,\ref{fig:schematic}\,(e)]. Note the increasing amplitude following each \code{prog1} pulse until saturation and the opposite behavior with each \code{prog2} pulse. This work used externally generated square programming pulses commensurate with hardware-in-the-loop training, but similar cells could be programmed directly with SPD pulses \cite{khan2023monolithic} for fully on-chip learning with local algorithms like spike-timing-dependent-plasticity.

\section{\label{sec:fab}Fabrication}
Circuits were fabricated at the NIST Boulder Microfabrication Facility in a 15-layer process. The full process details are described in the appendix of Ref.\,\onlinecite{khan2022superconducting}. The SPDs are made from MoSi and patterned into 200\,nm-wide meandering wires using electron-beam lithography \cite{verma2015high, lita2021mo}. We also use MoSi as a high kinetic inductance material for the larger-valued inductors (namely the synaptic integration inductor and the 22.5\,nH 8\,bit memory loop). The Josephson junctions are externally shunted Nb/a-Si/Nb tri-layers \cite{olaya2019planarized} with a target $I_\mathrm{c}$ of 100\,\textmu A. PdAu resistors were used for both the JJ shunts and as leak resistors in the synaptic integration loop. An integration time constant of 6.25\,\textmu s was targeted for all synapses.

\begin{figure}[!ht]
    \centering
\includegraphics[width=0.5\textwidth]{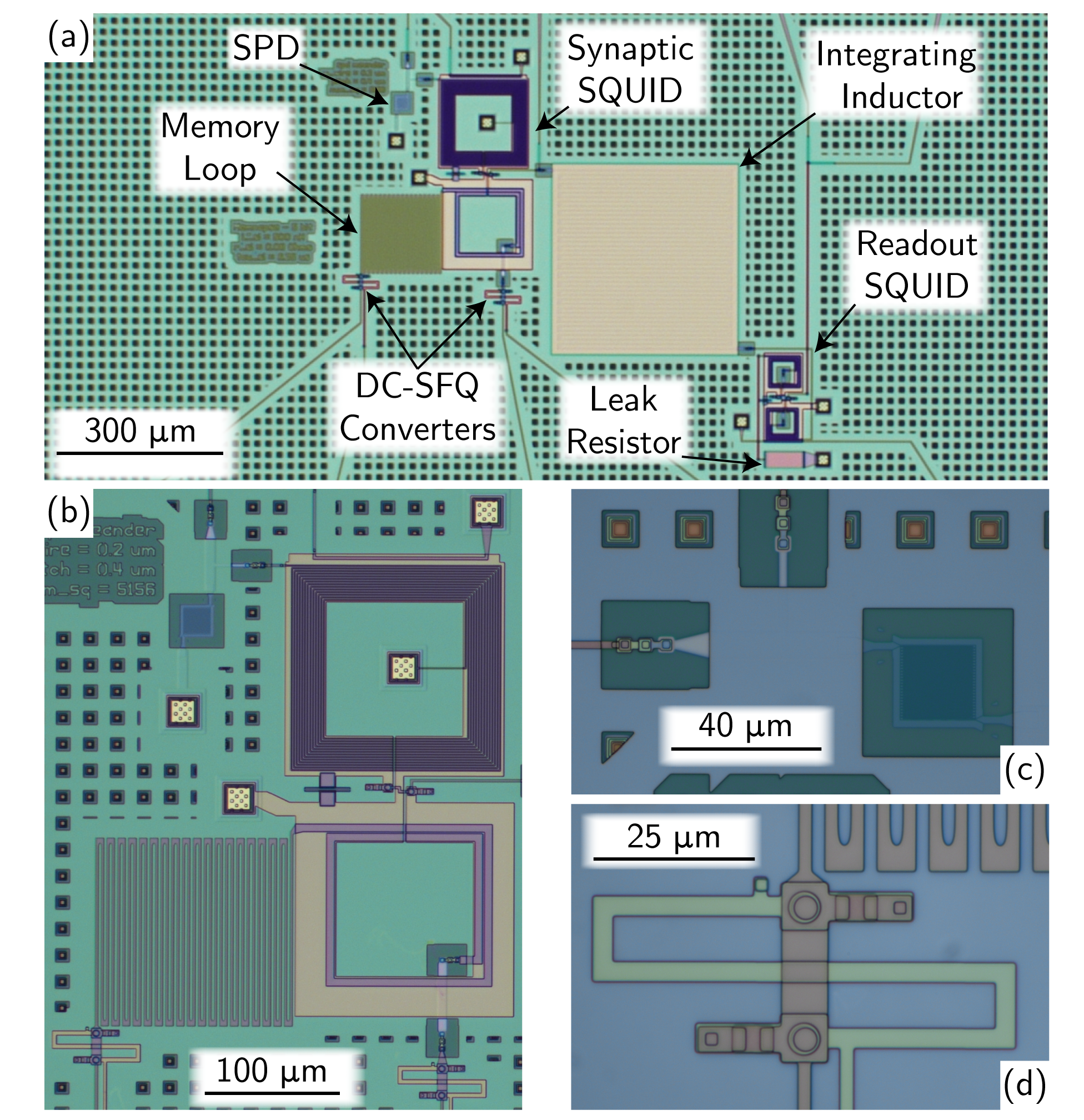}
    \caption{Microscope images of the 5\,bit synapse. (a) Full synapse. (b) Synaptic SQUID with SPD and memory loop inputs. (c) SPD (meander at right) along with vias and connections. (d) DC-SFQ converter for one side of the memory loop. JJs are the circles.}
    \label{fig:microscope}
\end{figure}

A microscope image of the full 5\,bit synapse is shown in Fig.\,\ref{fig:microscope}\,(a). The synaptic SQUID, SPD, and memory loop are shown in Fig.\,\ref{fig:microscope}\,(b). The synaptic SQUID uses a quadrupole configuration. This design both mitigates the effects of background magnetic fields varying over length-scales larger than the SQUID and provides a natural way to couple two independent input coils into the SQUID. The SPD drives the top input coil, while the memory loop drives the bottom. The SPD and a single DC-SFQ converter are shown in Figs.\,\ref{fig:microscope}\,(c) and (d). Both JJs are visible in Fig.\,\ref{fig:microscope}\,(d) (circles) along with external shunt resistors. 
No attempt was made to reduce the size of the circuits for these experiments. The full synapse is approximately 840\,\textmu m $\times$ 700\,\textmu m. Ref.\,\onlinecite{primavera2021considerations} estimates that similar synapses could be made as small as 30\,\textmu m $\times$ 30\,\textmu m in more advanced fabrication processes.

\section{\label{sec:data}Experimental Characterization}
\begin{figure*}[!ht]
    \centering
    \includegraphics{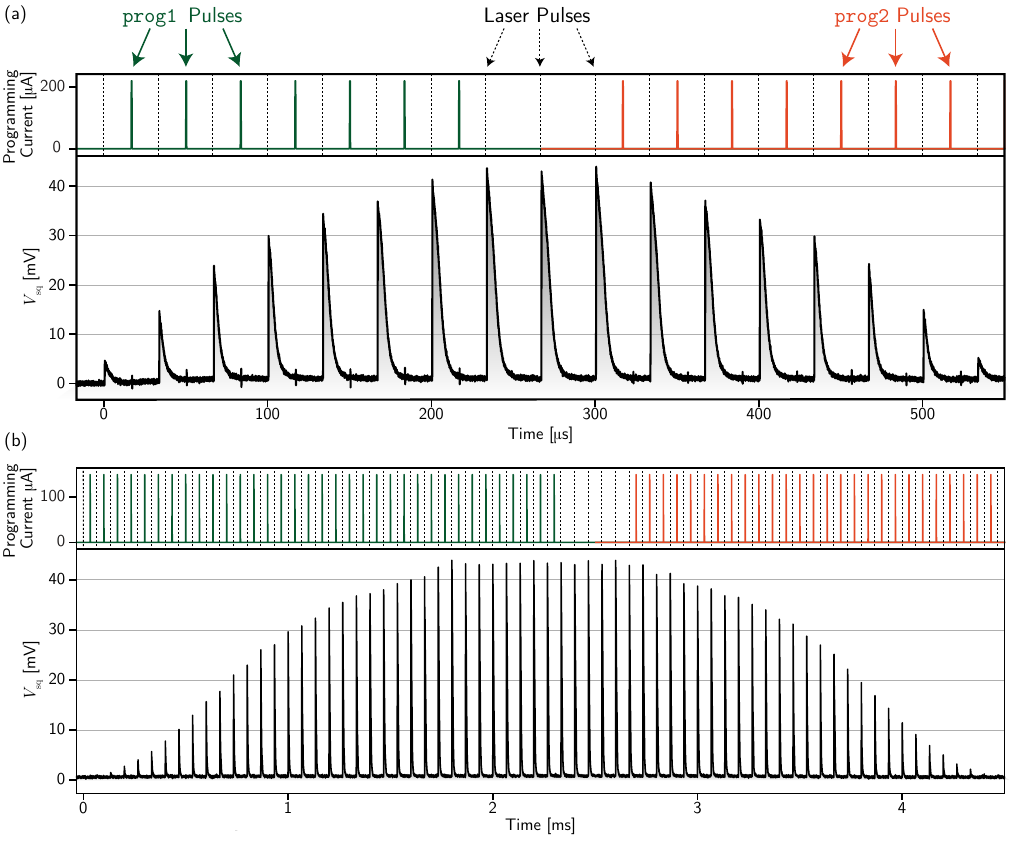}
    \caption{Synaptic response to individual laser pulses evolving with programming history. (a) 3\,bit synapse. Note how the three $V_\mathrm{sq}$ responses following the last \code{prog1} pulse maintain approximately the same peak value when no additional programming pulses are generated. (b) 5\,bit synapse. The post-synaptic response rises with each \code{prog1} pulse until the memory loop is saturated after about 28 pulses. The $V_\mathrm{sq}$ traces in both (a) and (b) are averaged 1000 times.}
    \label{fig:3-bit}
\end{figure*}

\begin{figure*}[!htb]
    \centering
    \includegraphics{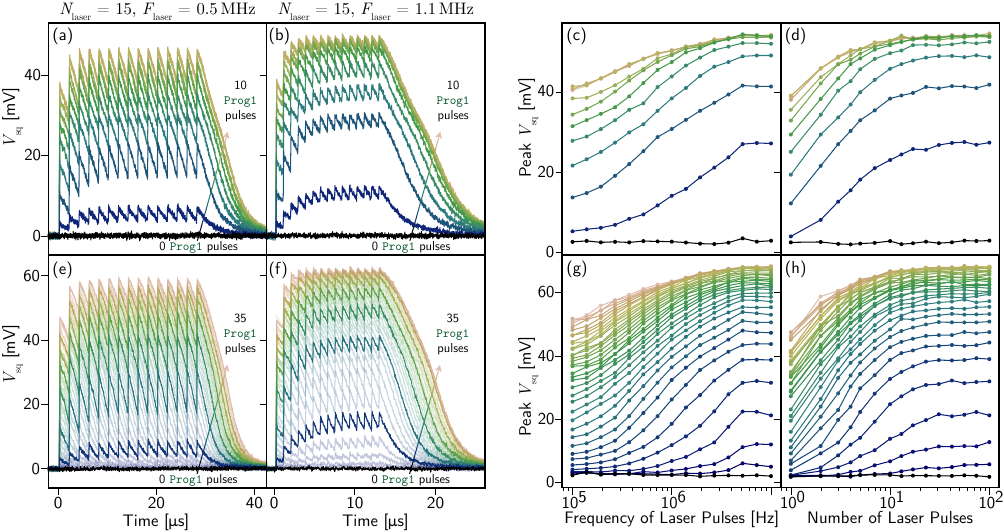}
    \caption{Integrating ability for the 3\,bit (a)-(d) and 5\,bit (e-h) synapses. (a) Response to 15 laser pulses arriving at 500\,kHz. Each trace corresponds to a different number of programming pulses sent into the memory loop at the beginning of the measurement. 500 averages. (b) Same as (a) except the laser pulse rate is increased to 1.1\,MHz. (c) Laser pulse frequency transfer function. The peak value of $V_\mathrm{sq}$ is plotted for different frequencies of optical pulses. The pulse number is fixed at 100 pulses. 200 averages. (d) Pulse number transfer function. The peak value of $V_\mathrm{sq}$ is plotted for number numbers of optical pulses. The pulse frequency is fixed at 10\,MHz. 200 averages. (e-h) repeats these plots for the 5\,bit synapse. In (e) and (f) every fifth trace is bolded for clarity.}
    \label{fig:Integrate}
\end{figure*}

\begin{figure*}
    \centering
    \includegraphics{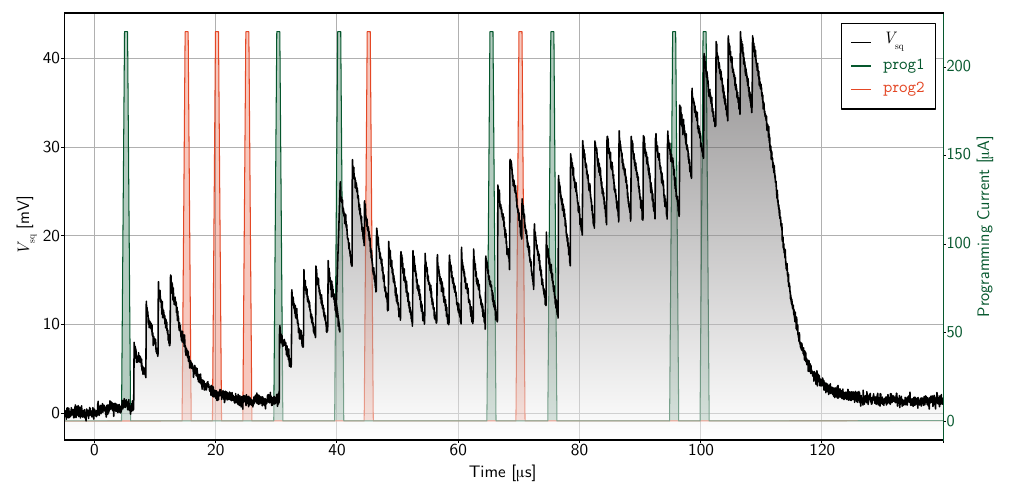}
    \caption{Dynamically changing the synaptic weight of the 3\,bit synapse while the synapse continues to receive and integrate signal. 500 averages. There is no observable cross-talk from the programming signals on the integrated synaptic current, allowing synaptic weights to be changed ``on-the-fly.''}
    \label{fig:Program}
\end{figure*}

Measurements were performed by inductively coupling the synaptic integration loops to another SQUID, which we refer to as the readout SQUID [Fig.\,\ref{fig:schematic}\,(a)]. The readout SQUID allows us to measure the voltage across the readout SQUID ($V_\mathrm{sq}$) as a proxy for $I_\mathrm{si}$. We note that $V_\mathrm{sq}$ is a somewhat distorted representation of $I_\mathrm{si}$, particularly when significant current is stored in the integration loop, due to the nonlinear response of the readout SQUID. Nonetheless, this convenient readout mechanism allows us to measure changes in $I_\mathrm{si}$ at sub-microsecond timescales and is essentially identical to how we envision coupling these synaptic signals into dendritic and somatic structures in future work \cite{primavera2021active, shainline2019fluxonic}. $V_\mathrm{sq}$ is amplified with a 60\,dB room temperature amplifier and recorded on a 1\,GHz oscilloscope. All plots report the amplified value of $V_\mathrm{sq}$.

All measurements were performed at 2.3\,K in a cryostat with a Gifford-McMahon cryocooler. The circuits were placed inside two concentric mu-metal shields to limit external magnetic noise. Nine coaxial cables were used for electrical input and output, and a single optical fiber was positioned above the test chip to flood-illuminate the entire sample. A 780\,nm laser with 480\,ps pulse width was used for optical input. The SPDs are not number-resolving detectors (except under special circumstances) and the laser pulse width is significantly shorter than the detector reset time ($\approx$30\,ns). Thus, even though the optical input is not in the single photon regime, the response of the SPD to a single laser pulse will not differ significantly from its single photon response. These responses are explored experimentally in Ref.\,\cite{buckley2020integrated} and in the supplement of Ref.\,\cite{khan2022superconducting}, where we operated the previous generation of synapses under low-light conditions and confirmed single-photon sensitivity. Additionally, averaging was performed on all time traces to counteract electrical noise obscuring the microvolt signals. The number of averages is given in each figure caption. While readout noise inhibits the ability to test the variance of the synaptic response, the amplitude of an averaged trace represents the mean of the underlying probability distribution of synaptic weight.

There are two additional current bias lines not shown in Fig.\,\ref{fig:schematic}\,(a) that are used to tune the the two SQUIDs. We refer to these as the ``addflux'' biases. The addflux lines couple flux into the SQUIDs to set the initial operating points (the zero flux point in Fig.\,\ref{fig:schematic}\,(b)). These operating points were tuned by hand before measurement began. In the future, on-chip magnetic shielding could be added to ensure that all SQUIDs begin with zero flux penetrating the loop. Additionally, every synapse was initialized in its lowest weight state by repeatedly pulsing the \code{prog2} port before each measurement.

For the first experiment, we demonstrate how the post-synaptic response to a single optical pulse evolves with the programming history of the memory loop. In Fig.\,\ref{fig:3-bit}\,(a), we initially alternate between pulsing the laser and applying programming pulses to \code{prog1} for the 3\,bit synapse. The post-synaptic current in the synaptic integration loop is allowed to decay to zero in between each successive laser pulse. We see that $V_\mathrm{sq}$ increases following each \code{prog1} programming pulse as desired. We then cease pulsing \code{prog1} and pulse the laser three times. As seen in the figure, these three pulses are nearly identical in height, confirming that the memory loop is indeed retaining its programmed state. We then begin pulsing \code{prog2} between laser pulses and see that the synaptic weight can be reduced one step at a time before the memory loop saturates at its lowest level after eight pulses. The experiment is repeated for the 5\,bit (620\,pH memory loop) synapse in Fig.\,\ref{fig:3-bit}\,(b). We witness the same qualitative behavior as with the 3\,bit synapse, but have significantly more states. \code{prog1} was pulsed 35 times in this experiment, although we observe that the post-synaptic height stops changing after about 28 pulses. This is due to the memory loop saturating at its maximum level slightly earlier than the designed 32 fluxons. It similarly takes about 28 \code{prog2} pulses to bring the synapse back to the minimal weight state, as expected. 

In Fig.\,\ref{fig:Integrate}, we demonstrate that the circuits also exhibit an integrating ability inspired by biological synapses \cite{magee2000dendritic}. If the laser is pulsed at a high enough frequency such that $I_\mathrm{si}$ does not decay fully between pulses, multiple detection events can be integrated over time. The time constant of the leak is determined by the L/R value of the synaptic integration loop. Although not the subject of this study, Ref.\,\cite{khan2022superconducting} demonstrates that synaptic integration times can be engineered across at least four orders of magnitude---hundreds of nanoseconds to several milliseconds. Fig.\,\ref{fig:Integrate}\,(a) shows the 3\,bit synapse responding to 15 laser pulses arriving at a 500\,kHz frequency. Each trace corresponds to a different initialization of the memory loop. Before the laser is turned on, the memory loop is given a fixed number of \code{prog1} pulses (0-10 in this case). We observe the synaptic response growing with each additional programming pulse until saturation at 8 pulses (the 8, 9, and 10 programming pulse traces are on top of each other). This experiment was repeated with a laser pulse frequency of 1.1\,MHz (b), where we observe a higher synaptic response for each programming condition than in (a), as expected from the leaky integrator.

In Figs.\,\ref{fig:Integrate}\,(c) and (d), the synaptic response is plotted as a function of the rate of incoming laser pulses (c) and the number of laser pulses in an input pulse train (d). In Fig.\,\ref{fig:Integrate}\,(c) the number of laser pulses is fixed at 100, but the frequency of those input pulses is swept from 100\,kHz to 10\,MHz. Each data point corresponds to the peak value of the synaptic response under those conditions (i.e. the peak value of a single trace of the type in (a) and (b)). Once again, each curve corresponds to a different weight initialization (0 - 10 \code{prog1} pulses). The demonstrated sensitivity to frequency is of particular interest in burst- and rate-coded applications. The complimentary measurement is presented in  Fig.\,\ref{fig:Integrate}\,(d), where the frequency of laser pulses is fixed at 10\,MHz, but the number of pulses is varied from 1 to 100. The eventual levelling off is characteristic of a leaky integrator reaching steady-state for sufficiently long pulse trains. In both (c) and (d), the ability of the memory loop to tune the synaptic response curve is evident. Figs.\,\ref{fig:Integrate} (e-h) show the same data for the 5\,bit synapse. In these plots, the number of \code{prog1} pulses is varied between 0 and 35 pulses. In (e) and (f), all 35 curves are plotted, but every fifth trace is shown with a darker line stroke for clarity. We see qualitatively similar data to the 3\,bit case, but with much higher synaptic weight resolution, as expected. 

The synapses presented here benefit from the ability of the DC-SFQ programming circuits to operate at timescales shorter than the synaptic integration dynamics. All experiments utilized programming pulse widths of approximately 100\,ns. This is significantly shorter than the 6.25\,\textmu s synaptic integration time and allows the synaptic weight to be changed dynamically, even while the synapse contains signal in its integration loop. This is illustrated in Fig.\,\ref{fig:Program} for the 3\,bit synapse. The laser is pulsed with a frequency of 700\,kHz. A series of programming pulses are input into the memory loop and we see the synaptic weight changing with every programming pulse (unless the loop is already saturated, as exhibited by the second and third \code{prog2} pulses). There is no observable cross-talk between the programming pulses and the integrated current, as the integration loop is isolated from the memory circuitry by the synaptic SQUID. This ability to change the weight dynamically is promising for future implementations of short term plasticity and homeostatic mechanisms. Further, memory updates can be completed in less time than the minimum inter-spike interval expected in SOENs hardware ($\approx 30$\,ns). Future implementations of on-chip learning in a network are therefore unlikely to ever be a bottleneck, even in the extreme case of a weight update after every synaptic event.

\begin{figure*}[!ht]
    \centering
    \includegraphics{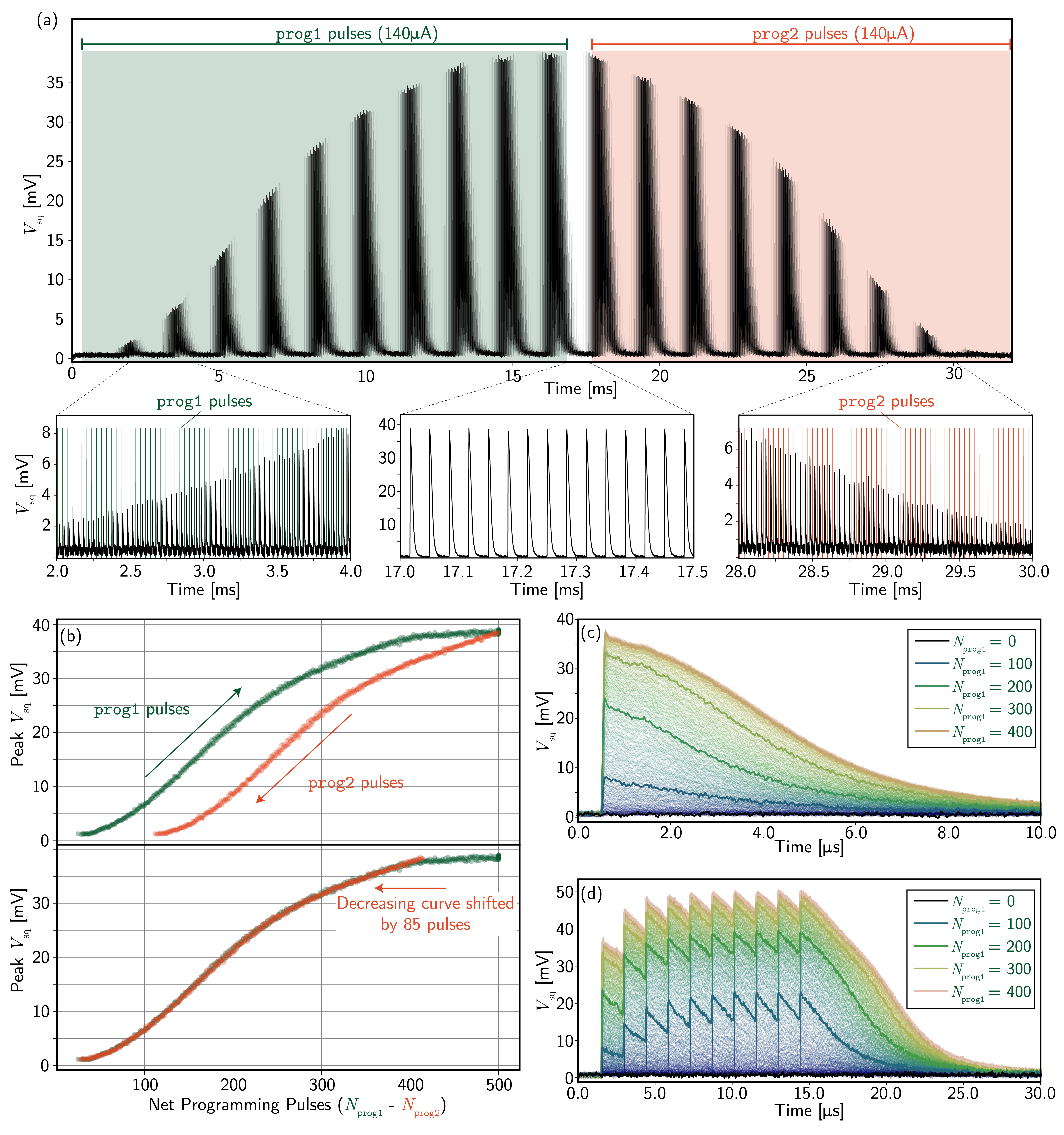}
    \caption{8\,bit device. (a) Alternating between laser and programming pulses, analogous to Fig.\,\ref{fig:3-bit}. Zoom-ins below resolve individual programming pulses and synaptic responses. Laser pulses are omitted for clarity. The \code{prog1} port is driven 500 times. 2500 averages. b) (Top) Peak $V_\mathrm{sq}$ as a function of the net number of programming pulses ($N_\mathrm{\code{prog1}} - N_\mathrm{\code{prog2}}$) applied to the memory loop. Green points are derived from the peaks while \code{prog1} is pulsed, while red points are derived from pulsing \code{prog2} following 500 \code{prog1} pulses. (Bottom) Same as above, except the red curve is shifted 85 pulses left. We see that the rising path and falling path are nearly symmetric. The 85 pulse offset implies that the last 85 \code{prog1} pulses occurred after the memory loop reached saturation and that there are approximately 415 states in the memory loop. (c) Response to a single laser pulse following different initializations of the memory loop (0 - 500 \code{prog1} pulses). The 0, 100, 200, 300, and 400 pulse traces are bolded for clarity. 1000 averages. (d) Same as (c) except responding to a pulse train of 10 laser pulses at 700 kHz.}
    \label{fig:8-bit}
\end{figure*}

In Fig.\,\ref{fig:8-bit}, we present the results of the synapse coupled to the largest memory loop ($\approx$22.5\,nH). We refer to this as the 8\,bit loop, although we estimate that there are actually over 400 distinct states. In the post-synaptic response, the difference between adjacent states is less than the noise floor of our measurement, but we can discern the same qualitative behavior as in the 3 and 5\,bit variants. In Fig.\,\ref{fig:8-bit}(a), we perform the same experiment as Fig.\,\ref{fig:3-bit}, alternating between individual laser and programming pulses. We pulse the potentiating port 500 times before reducing the synaptic weight back to zero. Zoomed-in portions show the synaptic weight increasing with the \code{prog1} pulses, remaining constant without any programming signals, and decreasing with \code{prog2} pulses. In Fig.\,\ref{fig:8-bit}\,(b) we plot the peak values of $V_\mathrm{sq}$ in part (a) following each laser pulse as a function of the net number of programming pulses, $N_\mathrm{\code{prog1}} - N_\mathrm{\code{prog2}}$. We break the plot into two parts: a green rising curve resulting from the region of \code{prog1} pulses, and a red decreasing curve from the later region of \code{prog2} pulses. These two curves are nearly symmetric, but shifted by 85 pulses as shown in the bottom panel of Fig.\,\ref{fig:8-bit}\,(b). Their symmetrical shapes suggest that the memory loop is indeed being programmed at the single fluxon level and both ports are operating correctly. The 85-pulse offset between the two curves implies that 85 of the \code{prog1} pulses came after the memory loop had already reached saturation. This allows us to estimate that the memory loop has a capacity of approximately $500 - 85 = 415$ states, or 8.7 bits. This is also in visual agreement with the apparent flattening of the green curve after about 415 \code{prog1} pulses. The slight upward slope in the region between 415 and 500 pulses is likely due to a small amount of integration of multiple laser pulses occurring at high synaptic weights. Figures \ref{fig:8-bit}\,(c) and (d) show the synapse responding to a single laser pulse and a 700\,kHz train of 10 laser pulses respectively. While noise obscures the small differences between states, the behavior is as expected.

\begin{figure}[!ht]
    \centering
    \includegraphics{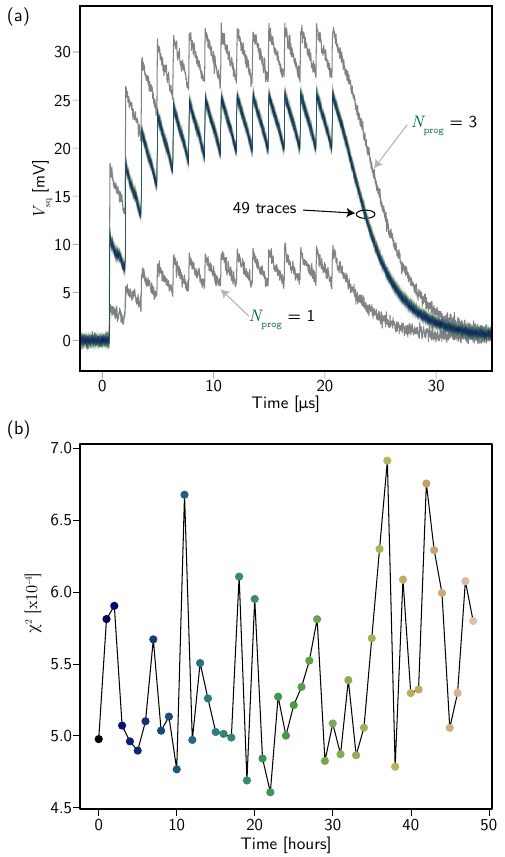}
    \caption{Stability of the 3\,bit synapse. (a) The grey traces correspond to the $N_\code{prog1} = 1$ and $N_\code{prog1} = 3$ states at the beginning of measurement period. All other traces correspond to the $N_\code{prog1} = 2$ state each taken one hour apart after a single initialization. 500 averages. (b) $\chi^2$ value (eq.\,\ref{eq:chi2}) of each trace as a function of time after initialization.}
    \label{fig:Stability}
\end{figure}
To quantify the volatility of the memory storage mechanism, the retention of one of the synapses was measured over a period of 48\,hours. In principle, the current stored in the memory loops should remain unchanged for as long as the circuit remains below the superconducting transition temperature. We test this on the 3\,bit synapse by first taking two reference traces with one and three \code{prog1} pulses. We then re-initialize the synapse with two \code{prog1} pulses. No other programming pulses were applied for the rest of the experiment. The synapse was driven once an hour with bursts of 15 laser pulses at a frequency of 700 kHz. Once again, averaging was required to reduce readout noise. 500 such bursts were applied (5 kHz burst frequency) and averaged to accurately measure the synaptic weight at each hour.  The results are presented in Fig.\,\ref{fig:Stability}\,(a). The two grey traces correspond to the $N_\code{prog1} = 1$ and $N_\code{prog1} = 3$ responses at the beginning of the experiment. There is no visible difference in the response over time and it never comes close to approaching either of the two adjacent states. In Fig.\,\ref{fig:Stability}\,(b), we plot the $\chi^2$ value of each trace as a function of time after programming. Eq.\,\ref{eq:chi2} is used to calculate the statistic.
\begin{equation}
    \chi^2(T) = \frac{\displaystyle\sum_{n=1}^{N_\mathrm{p}}(V_\mathrm{SQ,T}(n) - \overline{V_\mathrm{sq}}(n))^2}{\displaystyle\sum_{n=1}^{N_\mathrm{p}}\overline{V_\mathrm{sq}}(n)^2}
\label{eq:chi2}
\end{equation}
$V_\mathrm{SQ,T}(n)$ is the value of the $n^\mathrm{th}$ point in the trace taken $\mathrm{T}$ hours after programming, $N_\mathrm{p}$ is the number of data points in each trace, and $\overline{V_\mathrm{sq}}$ is the average of all 48 traces. There is no discernible trend over the 48\,hour measurement period and we expect that much longer retention times are possible. As a point of reference, we anticipate superconducting optoelectronic neurons to achieve spiking rates beyond 20\,MHz. Maximum spike rates in the brain are about 1\,kHz (for chattering neurons; most pyramidal neurons do not exceed gamma bursts of 80\,Hz). Using 20\,MHz/1\,kHz as a scale factor, 48 hours of stable memory in this hardware is roughly commensurate with 110 years of human brain activity.

\section{\label{sec:discussion}Discussion}
We have demonstrated programmable superconducting single-photon optoelectronic synapses and memory cells with over 8\,bit precision. For comparison, Intel's state-of-the-art  digital neuromorphic chip Loihi uses 9-bit precision \cite{davies2018loihi}. The programming energy for our synapses is approximately $2\Phi_0 I_c \approx $\,0.4\,aJ. Accounting for cooling would inflate this to around 0.4\,fJ. Assuming 1\% light-source efficiency, each presynaptic photon will require about 13\,fJ of energy. Programming energies are therefore unlikely to be a major contributor to the SOENs energy budget. Combined with potential programming speeds in excess of 100\,GHz, these synapses lend themselves to experimentation with sophisticated bio-inspired ``always-on'' plasticity mechanisms that would be too costly for other systems.  

While loop memory has many attractive features, it has long been considered a weakness of superconducting digital computing due to its low area density \cite{hilgenkamp2021josephson}. This is a fair criticism in the digital computing space, but much less so in SOENs neuromorphic hardware. Since every synapse already requires a SQUID, the addition of the memory loop is not a major contribution to the total area. While no effort was made to reduce area in this work, scaling analyses suggest that a single SOENs synapse with integrated memory should be realizable within a 30\,\textmu m$\times$30\,\textmu m area \cite{primavera2021considerations}. There is little motivation to shrink this area further due to the size of passive photonic components on other fabrication layers. Over one million neurons embedded in a network with a biologically-realistic average path length can still be expected to fit on a 300\,mm wafer \cite{primavera2021considerations}. While it is conceivable (although not at all obvious) that neuromorphic hardware of similar complexity based on silicon microelectronics could be more dense, the overall system scalability is far more limited due to communication bottlenecks. Further, superconducting loop memory outperforms emerging memrisitive technologies in all metrics other than area. While there is a wide range in reported performance of memristive devices \cite{zahoor2020resistive}, programming energies are typically on the order 100\,fJ/bit (more than two orders of magnitude greater than superconducting loop memory), write times are around 10\,ns (three orders of magnitude slower than superconducting memories), less than 6\,bit precision is typical \cite{kim2021multilevel}, and there are ongoing efforts to improve endurance and variability. Improved performance can be gained through more sophisticated programming protocols \cite{rao2023thousands}, but the programming time and cost of such schemes is best-suited for inference applications than large-scale training or online learning systems. Thus, where superconducting digital computing is weak relative to semiconductors with regard to memory, it appears to be strong in the domain of neural memory.

Although the demonstrated synapses are a marked improvement over the previous generation, there are several areas that require further research. First, future work with on-chip magnetic shielding and a cryostat with improved isolation could better elucidate the limits of bit precision in the devices. Second, there was no effort in this work to linearize the synaptic weighting. While the quantization of magnetic flux ensures that the stored current in each memory cell is nearly linear with its programming history, the amount of current added to the integration loop per detected photon is a function of the SQUID response, which is nonlinear. However, there has long been interest in improving the linearity of SQUID responses for sensing applications, and only slightly more complicated three-JJ SQUID designs have shown nearly linear responses \cite{kornev2009bi}. It may also be possible to design learning protocols where the nonlinearity does not pose an issue. Third, there is the key question of volatility. Superconducting loop memories do not require any power to maintain their state. Indeed, all the current biases to these synapses can be toggled off and on and the synapse will come back online in its previously programmed state. However, the temperature must remain below the critical temperature ($T_c$) of the superconducting materials (approximately 9\,K) or all information will be lost. There will likely be maintenance situations or power failures where the system must be warmed above $T_c$, and the ability to store the synaptic weights at room temperature would be beneficial. One possibility is to have a readout mechanism at every memory cell that could be used to measure the state of each memory cell and save this information digitally. The weight matrix could then be re-uploaded into the network once the system is cold again.

Local, programmable synaptic memory opens up a wide variety of potential applications. The present synapses could be useful in non-cognitive applications of spiking networks wherein a weight-adjacency matrix representing a specific problem is programmed into the network during an initialization phase. The network behavior is then allowed to evolve in time, with the resulting network dynamics encoding the solution to the problem. This scheme has been used to solve optimization problems and systems of differential equations \cite{aimone2022review}. In terms of artificial intelligence, the present synapses are already well-suited for inference applications utilizing a matrix of pre-learned weights. They are also amenable to hardware-in-the-loop training, in which a standard digital computer generates the appropriate programming pulses from observing the network's output and internal variables. While such training could be performed at a small scale with memory-less synapses, this scheme would require an independent current bias for every weight in the network, which is infeasible for large systems due to the heat load incurred by electrical lines. In contrast, these programmable synapses could be interfaced with cryo-CMOS control circuitry and an addressing system for large-scale programming using a limited number of lines between the cryostat and room temperature. The ultimate goal, however, is to remove the digital computer from the training loop. This will require the development of learning algorithms and plasticity circuits to update the synaptic memories using primarily local information. We have developed a simulation framework for SOENs hardware \cite{shainline2023phenomenological} and recently demonstrated a neural network utilizing this synapse design to solve simple image classification problems with local updates in conjunction with a global reward signal \cite{Ryan2023ICONS}. Additionally, plasticity circuits implementing a spike-timing-dependent plasticity update rule with very similar memory elements were presented in Ref.\,\cite{shainline2019superconducting}. Such plasticity circuits must receive optical, rather electrical inputs, since the native spiking behavior in SOENs is performed in the optical domain. The single-photon-to-single-fluxon converters presented in Ref\,\cite{khan2023monolithic} are an important step in this direction and can be considered optically programmable memory cells in their own right. Future work will focus on developing these plasticity circuits, investigating more sophisticated dendritic processing \cite{shainline2019fluxonic}, and demonstrating full superconducting optoelectronic neurons by integrating these synapses with on-chip light sources and passive integrated-photonic interconnection networks.

\section{Acknowledgements}
We thank Dr. Michael Schneider for useful feedback on the manuscript. This work was made possible by the institutional support from the National Institute of Standards and Technology (award no. 70NANB18H006).  

\bibliographystyle{unsrtnat}
\bibliography{biblio}

\begin{thebibliography}{28}
\providecommand{\natexlab}[1]{#1}
\providecommand{\url}[1]{\texttt{#1}}
\expandafter\ifx\csname urlstyle\endcsname\relax
  \providecommand{\doi}[1]{doi: #1}\else
  \providecommand{\doi}{doi: \begingroup \urlstyle{rm}\Url}\fi

\bibitem[Sebastian et~al.(2020)Sebastian, Le~Gallo, Khaddam-Aljameh, and Eleftheriou]{sebastian2020memory}
Abu Sebastian, Manuel Le~Gallo, Riduan Khaddam-Aljameh, and Evangelos Eleftheriou.
\newblock Memory devices and applications for in-memory computing.
\newblock \emph{Nature nanotechnology}, 15\penalty0 (7):\penalty0 529--544, 2020.

\bibitem[Shainline et~al.(2019)Shainline, Buckley, McCaughan, Chiles, Jafari~Salim, Castellanos-Beltran, Donnelly, Schneider, Mirin, and Nam]{shainline2019superconducting}
Jeffrey~M Shainline, Sonia~M Buckley, Adam~N McCaughan, Jeffrey~T Chiles, Amir Jafari~Salim, Manuel Castellanos-Beltran, Christine~A Donnelly, Michael~L Schneider, Richard~P Mirin, and Sae~Woo Nam.
\newblock Superconducting optoelectronic loop neurons.
\newblock \emph{Journal of Applied Physics}, 126\penalty0 (4):\penalty0 044902, 2019.

\bibitem[Shainline(2021)]{shainline2021optoelectronic}
Jeffrey~M Shainline.
\newblock Optoelectronic intelligence.
\newblock \emph{Applied Physics Letters}, 118\penalty0 (16), 2021.

\bibitem[Khan et~al.(2022)Khan, Primavera, Chiles, McCaughan, Buckley, Tait, Lita, Biesecker, Fox, Olaya, et~al.]{khan2022superconducting}
Saeed Khan, Bryce~A Primavera, Jeff Chiles, Adam~N McCaughan, Sonia~M Buckley, Alexander~N Tait, Adriana Lita, John Biesecker, Anna Fox, David Olaya, et~al.
\newblock Superconducting optoelectronic single-photon synapses.
\newblock \emph{Nature Electronics}, 5\penalty0 (10):\penalty0 650--659, 2022.

\bibitem[Burr et~al.(2017)Burr, Shelby, Sebastian, Kim, Kim, Sidler, Virwani, Ishii, Narayanan, Fumarola, et~al.]{burr2017neuromorphic}
Geoffrey~W Burr, Robert~M Shelby, Abu Sebastian, Sangbum Kim, Seyoung Kim, Severin Sidler, Kumar Virwani, Masatoshi Ishii, Pritish Narayanan, Alessandro Fumarola, et~al.
\newblock Neuromorphic computing using non-volatile memory.
\newblock \emph{Advances in Physics: X}, 2\penalty0 (1):\penalty0 89--124, 2017.

\bibitem[Crowe(1957)]{crowe1957trapped}
James~W Crowe.
\newblock Trapped-flux superconducting memory.
\newblock \emph{IBM Journal of Research and Development}, 1\penalty0 (4):\penalty0 294--303, 1957.

\bibitem[Tahara and Wada(1987)]{tahara1987vortex}
Shuichi Tahara and Yosifusa Wada.
\newblock A vortex transitional ndro josephson memory cell.
\newblock \emph{Japanese journal of applied physics}, 26\penalty0 (9R):\penalty0 1463, 1987.

\bibitem[Schegolev et~al.(2020)Schegolev, Klenov, Soloviev, and Tereshonok]{schegolev2020learning}
Andrey Schegolev, Nikolay Klenov, Igor Soloviev, and Maxim Tereshonok.
\newblock Learning cell for superconducting neural networks.
\newblock \emph{Superconductor Science and Technology}, 34\penalty0 (1):\penalty0 015006, 2020.

\bibitem[Primavera and Shainline(2021{\natexlab{a}})]{primavera2021considerations}
Bryce~A Primavera and Jeffrey~M Shainline.
\newblock Considerations for neuromorphic supercomputing in semiconducting and superconducting optoelectronic hardware.
\newblock \emph{Frontiers in Neuroscience}, 15:\penalty0 732368, 2021{\natexlab{a}}.

\bibitem[Duzer and Turner(1998)]{van1998principles}
T.~Van Duzer and C.W. Turner.
\newblock \emph{Principles of superconductive devices and circuits}.
\newblock Prentice {H}all, USA, second edition, 1998.

\bibitem[Kadin(1999)]{kadin1999introduction}
Alan~M Kadin.
\newblock \emph{Introduction to superconducting circuits}.
\newblock Wiley-Interscience, 1999.

\bibitem[Khan et~al.(2023)Khan, Primavera, Mirin, Nam, and Shainline]{khan2023monolithic}
Saeed Khan, Bryce~A Primavera, Richard~P Mirin, Sae~Woo Nam, and Jeffrey~M Shainline.
\newblock Monolithic integration of superconducting-nanowire single-photon detectors with josephson junctions for scalable single-photon sensing.
\newblock \emph{arXiv preprint arXiv:2310.13107}, 2023.

\bibitem[Verma et~al.(2015)Verma, Korzh, Bussieres, Horansky, Dyer, Lita, Vayshenker, Marsili, Shaw, Zbinden, et~al.]{verma2015high}
Varun~B Verma, Boris Korzh, Felix Bussieres, Robert~D Horansky, Shellee~D Dyer, Adriana~E Lita, Igor Vayshenker, Francesco Marsili, Matthew~D Shaw, Hugo Zbinden, et~al.
\newblock High-efficiency superconducting nanowire single-photon detectors fabricated from mosi thin-films.
\newblock \emph{Optics express}, 23\penalty0 (26):\penalty0 33792--33801, 2015.

\bibitem[Lita et~al.(2021)Lita, Verma, Chiles, Mirin, and Nam]{lita2021mo}
Adriana~E Lita, Varun~B Verma, Jeff Chiles, Richard~P Mirin, and Sae~Woo Nam.
\newblock Mo x si1- xa versatile material for nanowire to microwire single-photon detectors from uv to near ir.
\newblock \emph{Superconductor Science and Technology}, 34\penalty0 (5):\penalty0 054001, 2021.

\bibitem[Olaya et~al.(2019)Olaya, Castellanos-Beltran, Pulecio, Biesecker, Khadem, Lewitt, Hopkins, Dresselhaus, and Benz]{olaya2019planarized}
David Olaya, Manuel Castellanos-Beltran, Javier Pulecio, John Biesecker, Soroush Khadem, Theodore Lewitt, Peter Hopkins, Paul Dresselhaus, and Samuel Benz.
\newblock Planarized process for single-flux-quantum circuits with self-shunted nb/nb $ \_ $\{$x$\}$ $ si $ \_ $\{$1-x$\}$ $/nb josephson junctions.
\newblock \emph{IEEE Transactions on Applied Superconductivity}, 29\penalty0 (6):\penalty0 1--8, 2019.

\bibitem[Primavera and Shainline(2021{\natexlab{b}})]{primavera2021active}
Bryce~A Primavera and Jeffrey~M Shainline.
\newblock An active dendritic tree can mitigate fan-in limitations in superconducting neurons.
\newblock \emph{Applied Physics Letters}, 119\penalty0 (24):\penalty0 242601, 2021{\natexlab{b}}.

\bibitem[Shainline(2019)]{shainline2019fluxonic}
Jeffrey~M Shainline.
\newblock Fluxonic processing of photonic synapse events.
\newblock \emph{IEEE Journal of Selected Topics in Quantum Electronics}, 26\penalty0 (1):\penalty0 1--15, 2019.

\bibitem[Buckley et~al.(2020)Buckley, Tait, Chiles, McCaughan, Khan, Mirin, Nam, and Shainline]{buckley2020integrated}
Sonia~M Buckley, Alexander~N Tait, Jeffrey Chiles, Adam~N McCaughan, Saeed Khan, Richard~P Mirin, Sae~Woo Nam, and Jeffrey~M Shainline.
\newblock Integrated-photonic characterization of single-photon detectors for use in neuromorphic synapses.
\newblock \emph{Physical Review Applied}, 14\penalty0 (5):\penalty0 054008, 2020.

\bibitem[Magee(2000)]{magee2000dendritic}
Jeffrey~C Magee.
\newblock Dendritic integration of excitatory synaptic input.
\newblock \emph{Nature Reviews Neuroscience}, 1\penalty0 (3):\penalty0 181--190, 2000.

\bibitem[Davies et~al.(2018)Davies, Srinivasa, Lin, Chinya, Cao, Choday, Dimou, Joshi, Imam, Jain, et~al.]{davies2018loihi}
Mike Davies, Narayan Srinivasa, Tsung-Han Lin, Gautham Chinya, Yongqiang Cao, Sri~Harsha Choday, Georgios Dimou, Prasad Joshi, Nabil Imam, Shweta Jain, et~al.
\newblock Loihi: A neuromorphic manycore processor with on-chip learning.
\newblock \emph{Ieee Micro}, 38\penalty0 (1):\penalty0 82--99, 2018.

\bibitem[Hilgenkamp(2021)]{hilgenkamp2021josephson}
Hans Hilgenkamp.
\newblock Josephson memories.
\newblock \emph{Journal of superconductivity and novel magnetism}, 34\penalty0 (6):\penalty0 1621--1625, 2021.

\bibitem[Zahoor et~al.(2020)Zahoor, Azni~Zulkifli, and Khanday]{zahoor2020resistive}
Furqan Zahoor, Tun~Zainal Azni~Zulkifli, and Farooq~Ahmad Khanday.
\newblock Resistive random access memory (rram): an overview of materials, switching mechanism, performance, multilevel cell (mlc) storage, modeling, and applications.
\newblock \emph{Nanoscale research letters}, 15:\penalty0 1--26, 2020.

\bibitem[Kim et~al.(2021)Kim, Kim, Hong, Park, Hwang, Park, and Kim]{kim2021multilevel}
Tae-Hyeon Kim, Sungjoon Kim, Kyungho Hong, Jinwoo Park, Yeongjin Hwang, Byung-Gook Park, and Hyungjin Kim.
\newblock Multilevel switching memristor by compliance current adjustment for off-chip training of neuromorphic system.
\newblock \emph{Chaos, Solitons \& Fractals}, 153:\penalty0 111587, 2021.

\bibitem[Rao et~al.(2023)Rao, Tang, Wu, Song, Zhang, Yin, Zhuo, Kiani, Chen, Jiang, et~al.]{rao2023thousands}
Mingyi Rao, Hao Tang, Jiangbin Wu, Wenhao Song, Max Zhang, Wenbo Yin, Ye~Zhuo, Fatemeh Kiani, Benjamin Chen, Xiangqi Jiang, et~al.
\newblock Thousands of conductance levels in memristors integrated on cmos.
\newblock \emph{Nature}, 615\penalty0 (7954):\penalty0 823--829, 2023.

\bibitem[Kornev et~al.(2009)Kornev, Soloviev, Klenov, and Mukhanov]{kornev2009bi}
VK~Kornev, II~Soloviev, NV~Klenov, and OA~Mukhanov.
\newblock Bi-squid: a novel linearization method for dc squid voltage response.
\newblock \emph{Superconductor Science and Technology}, 22\penalty0 (11):\penalty0 114011, 2009.

\bibitem[Aimone et~al.(2022)Aimone, Date, Fonseca-Guerra, Hamilton, Henke, Kay, Kenyon, Kulkarni, Mniszewski, Parsa, et~al.]{aimone2022review}
James Aimone, Prasanna Date, Gabriel Fonseca-Guerra, Kathleen Hamilton, Kyle Henke, Bill Kay, Garrett Kenyon, Shruti Kulkarni, Susan Mniszewski, Maryam Parsa, et~al.
\newblock A review of non-cognitive applications for neuromorphic computing.
\newblock \emph{Neuromorphic Computing and Engineering}, 2022.

\bibitem[Shainline et~al.(2023)Shainline, Primavera, and Khan]{shainline2023phenomenological}
Jeffrey~M Shainline, Bryce~A Primavera, and Saeed Khan.
\newblock Phenomenological model of superconducting optoelectronic loop neurons.
\newblock \emph{Physical Review Research}, 5\penalty0 (1):\penalty0 013164, 2023.

\bibitem[O'Loughlin et~al.(2023)O'Loughlin, Primavera, and Shainline]{Ryan2023ICONS}
Ryan O'Loughlin, Bryce Primavera, and Jeffrey Shainline.
\newblock Dendritic learning in superconducting optoelectronic networks.
\newblock In \emph{Proceedings of the 2023 International Conference on Neuromorphic Systems}, ICONS '23, New York, NY, USA, 2023. Association for Computing Machinery.
\newblock ISBN 9798400701757.
\newblock \doi{10.1145/3589737.3605972}.

\end{thebibliography}

\appendix
\section*{Appendices}
\addcontentsline{toc}{section}{Appendices}
\renewcommand{\thesubsection}{\Alph{subsection}}

\subsection{Further Experimental Details}
\begin{figure*}[!ht]
    \centering
    \includegraphics{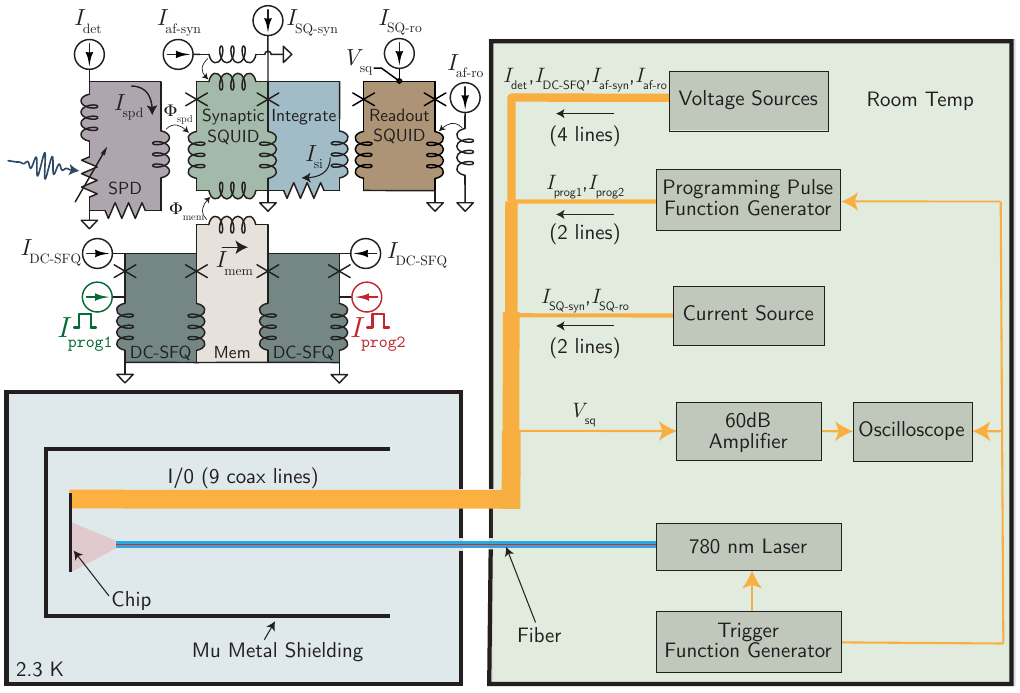}
    \caption{Experimental Set-Up. Room temperature electronics in the green box control cryogenic circuits inside the fridge (blue).}
    \label{fig:apx_setup}
\end{figure*}

Figure \ref{fig:apx_setup} provides a schematic of the full experimental set-up. One function generator (bottom) is used to generate a synchronization signal that is sent to the laser, the oscilloscope, and a second two-channel function generator for the programming signals. $I_\mathrm{SQ-syn}$ and $I_\mathrm{SQ-ro}$ are generated with a commercially available DC current source. All other DC biases ($I_\mathrm{spd}$, $I_\mathrm{DC-SFQ}$, $I_\mathrm{af-syn}$, and $I_\mathrm{af-ro}$) are generated with bias resistors in series with DC voltage sources.

Finding the appropriate biases for each synapse before data collection begins with a sweep of $I_\mathrm{af-ro}$ at several different values of $I_\mathrm{SQ-ro}$. The value of $I_\mathrm{SQ-ro}$ that maximizes the difference between the peak and the trough of the readout SQUID response was chosen as the readout SQUID bias for all subsequent measurements. $I_\mathrm{af-ro}$ would then be set to minimize $V_\mathrm{sq}$, providing the maximum dynamic range for measurements of $I_\mathrm{si}$. The SPDs had good uniformity and $I_\mathrm{spd}$ was set to 14\,\textmu A for all experiments. The rest of the biases had to be set with an iterative procedure by hand to maximize the number of states while keeping the memory loop operating in the DC-SFQ regime and ensuring that the synaptic SQUID was confined to only a single period of the SQUID response. Future circuits fabricated in mature superconducting foundries with more uniform JJ $I_c$ and additional superconducting layers for magnetic shielding should require significantly less hand-tuning. We also note that the memory loops themselves may be used as tunable elements to offset background flux and some variation in $I_c$ in future calibration procedures.

\begin{figure*}[!ht]
    \centering
    \includegraphics{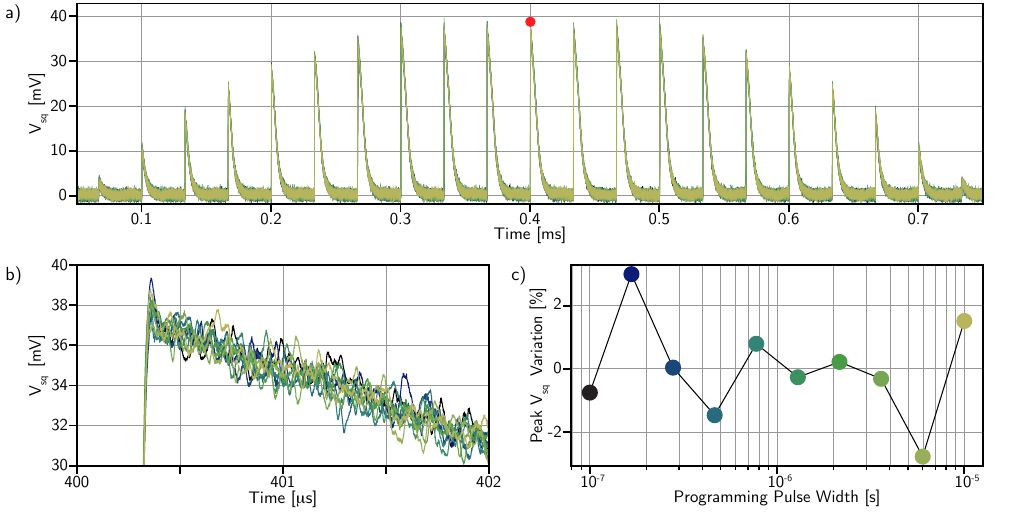}
    \caption{Programming pulse width sweep confirming single fluxon operation. (a) The 3\,bit synaptic response to individual programming pulses (not shown) similar to Fig.\,\ref{fig:3-bit}\,(a). The curves of 10 different programming pulse widths (100\,ns - 10\,\textmu s) are plotted on top of each other. 200 averages. (b) Zoom-in of peak indicated with red dot in (a). There is no discernible difference between the traces. (c) Variation of the peak values of each trace in (b) as a function of the programming width. Y-axis gives the variation of each data point with respect to the mean of all ten points. There is no trend with respect to programming pulse width, as expected for properly functioning DC-SFQ converters.}
    \label{fig:Pw_Sweep}
\end{figure*}

In Fig.\,\ref{fig:Pw_Sweep} we confirm the single fluxon operation of the DC-SFQ converters. The experiment measuring the single optical pulse response as a function of programming history (Fig.\,\ref{fig:3-bit}) was repeated for 10 different programming pulse widths. These pulse widths were geometrically spaced from 100\,ns to 10\,\textmu s. If DC-SFQ converters are operating properly, there should be no dependence on the programming pulse width. In Fig.\,\ref{fig:Pw_Sweep}\,(a) the traces for each pulse width are plotted on top of each other. Zooming in on one peak (b) and plotting the variation in height as a function of pulse width (c) confirms proper operation over at least two orders of magnitude in pulse width. This is not surprising; if the programming circuitry was not operating in the DC-SFQ regime, the memory loops would have saturated for pulse widths larger than about 100\,ps due to the small loop inductance.

\subsection{Synapse Response Maps}
\begin{figure*}[!ht]
    \centering
    \includegraphics{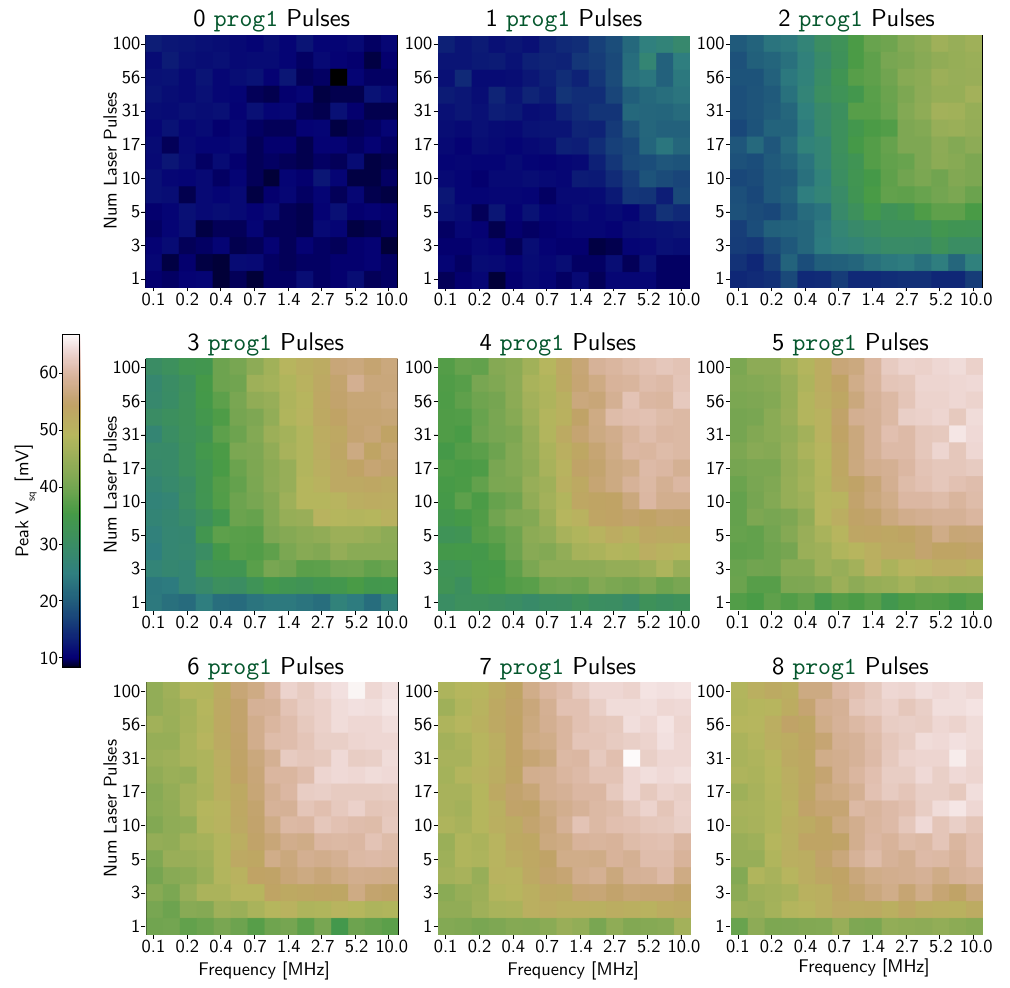}
    \caption{3\,bit synaptic response. The peak value of $V_\mathrm{sq}$ is recorded for optical pulse trains of varying frequency (x-axes) and varying pulse number (y-axes). Each of the nine squares differs from the others only by the number of \code{prog1} programming pulses applied before the laser is turned on. 25 averages.}
    \label{fig:3bit_heat_map}
\end{figure*}

\begin{figure*}[!ht]
    \centering
    \includegraphics{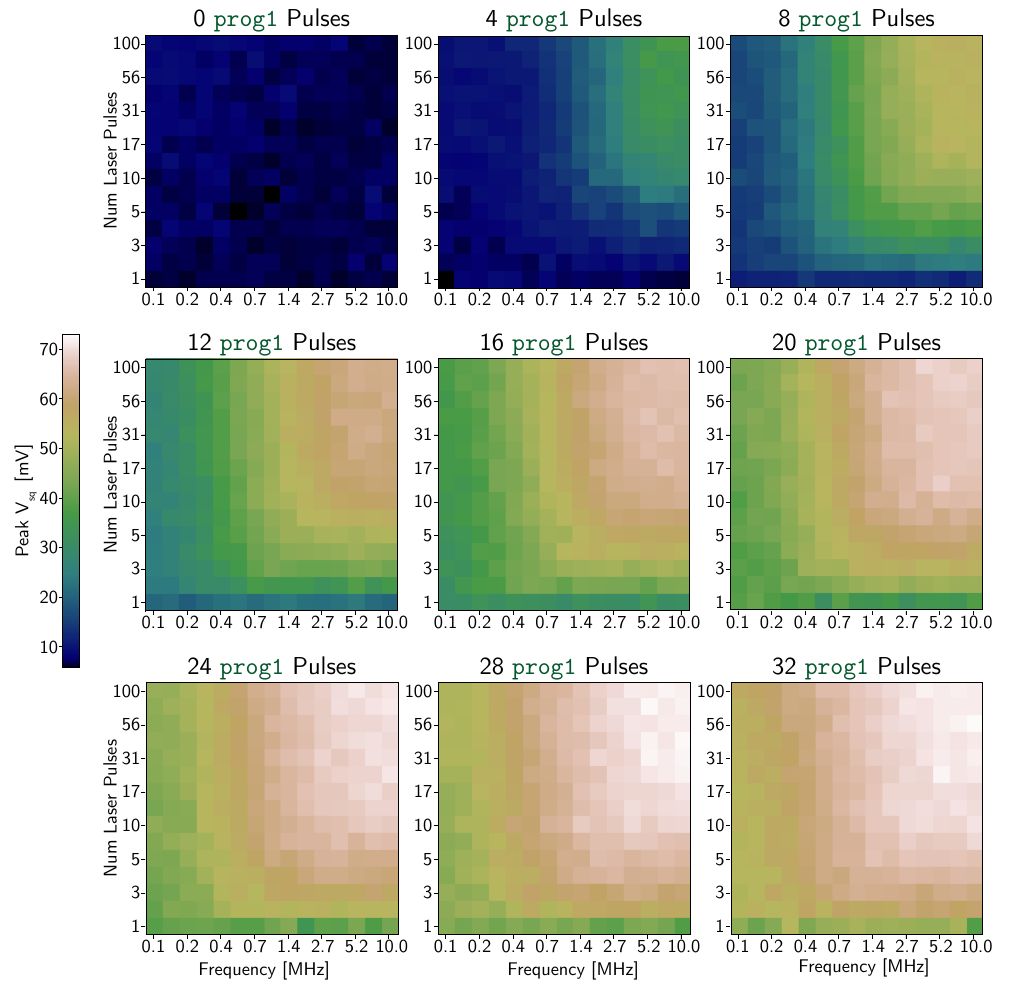}
    \caption{5\,bit synaptic response. The peak value of $V_\mathrm{sq}$ is recorded for optical pulse trains of varying frequency (x-axes) and varying pulse number (y-axes). 50 averages. Compare with 3\,bit data in Fig.\,\ref{fig:3bit_heat_map}. Both synapses exhibit very similar behavior, but the effect of individual programming pulses is greatly reduced in the 5\,bit case.}
    \label{fig:5bit_heat_map}
\end{figure*}

The 3\,bit and 5\,bit synapses were tested under many different conditions to confirm correct operation. In Fig.\,\ref{fig:3bit_heat_map} and Fig.\,\ref{fig:5bit_heat_map}, each of the nine heat maps corresponds to a given initialized synaptic weight (a certain number of \code{prog1} pulses added to the memory loop before the laser is turned on). Each pixel in each heat map corresponds to the maximum value of $V_\mathrm{sq}$ that was recorded for a specific sequence of optical pulses. These plots are extensions of the transfer functions in Fig.\,\ref{fig:Integrate} across a wider parameter space. Figure 5 in Ref.\,\cite{khan2022superconducting} is a useful comparison to these plots. That work demonstrated the effects of changing the parameters of the integration loop, but used external current biases to establish synaptic weights. This work shows that similar results can be achieved with local programmable memory elements.

\end{document}